\documentclass[11pt,english]{article}

\usepackage{authblk}
\usepackage{amsthm,amsmath}
\RequirePackage[square,numbers]{natbib}
\usepackage{bm}
\usepackage{graphicx}

\begin{document}

\title{DM-PhyClus: A Bayesian phylogenetic algorithm for infectious disease transmission cluster inference}

\author[1]{Luc Villandr\'{e}\thanks{\textit{Corresponding author}: luc.villandre@mail.mcgill.ca}}
\affil[1]{Dept. of Epidemiology, Biostatistics, and Occ. Health\\
	  McGill University, Montreal, QC, Canada}

\author[2]{Aur\'{e}lie Labbe}
\affil[2]{Dept. of Decision Sciences\\
	  HEC Montr\'{e}al, Montreal, QC, Canada}
	  
\author[3]{Bluma Brenner}
\affil[3]{McGill AIDS Centre\\
	  Lady Davis Institute, Jewish General Hospital, Montreal, QC, Canada}
	  
\author[4,5]{Michel Roger}
\affil[4]{Centre de Recherche du Centre Hospitalier de l'Universit\'{e} ́de Montr\'{e}al (CRCHUM)\\
  Montreal, QC, Canada}
\affil[5]{D\'{e}partement de microbiologie, infectiologie et immunologie\\
	  Universit\'{e} de Montr\'{e}al, Montreal, QC, Canada}
	  
\author[6]{David A. Stephens}
\affil[6]{Dept. of Mathematics and Statistics\\
	  McGill University, Montreal, QC, Canada}
% \author{ 
%   McGill University\\
%   Dept. of Epidemiology, Biostatistics, and Occ. Health\\
%   \texttt{luc.villandre@mail.mcgill.ca}
%   \and
%   Labbe, Aur\'{e}lie \\
%   \texttt{aurelie.labbe@hec.ca}
% }
% \affil{
%   Department of Epidemiology, Biostatistics, and Occupational Health, McGill University
%   1020 avenue des Pins Ouest
%   Montreal, QC, Canada
%   H3A 1A2  
% }
% \email{luc.villandre@mail.mcgill.ca}
\maketitle

\begin{abstract}
\textbf{Background.} Conventional phylogenetic clustering approaches rely on arbitrary cutpoints applied a posteriori to phylogenetic estimates. Although in practice, Bayesian and bootstrap-based clustering tend to lead to similar estimates, they often produce conflicting measures of confidence in clusters. The current study proposes a new Bayesian phylogenetic clustering algorithm, which we refer to as \textit{DM-PhyClus}, that identifies sets of sequences resulting from quick transmission chains, thus yielding easily-interpretable clusters, without using any ad hoc distance or confidence requirement. 
\textbf{Results.} Simulations reveal that DM-PhyClus can outperform conventional clustering methods, as well as the Gap procedure, a pure distance-based algorithm, in terms of mean cluster recovery. We apply DM-PhyClus to a sample of real HIV-1 sequences, producing a set of clusters whose inference is in line with the conclusions of a previous thorough analysis.
\textbf{Conclusions.} DM-PhyClus, by eliminating the need for cutpoints and producing sensible inference for cluster configurations, can facilitate transmission cluster detection. Future efforts to reduce incidence of infectious diseases, like HIV-1, will need reliable estimates of transmission clusters. It follows that algorithms like DM-PhyClus could serve to better inform public health strategies.
\end{abstract}

\section*{Introduction}

The collection and, often public, availability of viral genotyping data has made phylogenetics, the field concerned with the inference from genetic data of the ancestral history of organisms, a popular tool for modelling epidemics \cite{Foley2013, Huerta-Cepas2014}. Phylogenetic models represent the ancestral relationships between sequences of nucleotides or amino acids with a hierarchical tree structure known as a \textit{phylogeny}. Phylogenetics can help guide public health efforts to curb incidence of HIV-1 and tuberculosis \cite{Brenner2013, Brenner2013a, VanderSpoelvanDijk2016}, by revealing the existence of \textit{transmission clusters}, epidemiologically-linked individuals infected by a genetically-similar pathogen. Transmission clusters are known to affect incidence and may hinder the implementation of effective intervention strategies \cite{Brenner2011a}.

\subsection*{Transmission cluster inference}

Observed clustering in viral sequencing data, thought to result from series of fast onward transmission events called \textit{quick transmission chains}, is a convenient proxy for transmission clusters \cite{Brenner2007}. To estimate transmission clusters from an inferred phylogeny, a collection of ad hoc rules are conventionally applied. One normally looks for a partition of the sample into \textit{clades}. A clade is a set of sequences corresponding to all tips descended from a given ancestral node in the tree. Usually, a clade corresponds to a cluster only when it is known with high confidence, and when its sequences are similar. Unsurprisingly, disagreements over clustering rules are common, and what the resulting partitions mean in an epidemiological sense is still unclear \cite{Chalmet2010, Villandre2016}. 

\subsection*{Study objective}

In the present study, we aim to propose a new Bayesian phylogenetic clustering algorithm, called \textit{DM-PhyClus}, that eliminates the need for arbitrary distance and confidence criteria. DM-PhyClus looks directly for sets of sequences resulting from quick transmission chains, thus also improving interpretability of clusters.

\subsection*{Phylogenetic inference and clustering}

Bayesian phylogenetic inference is commonly used in the clustering of sequencing data, mainly because it readily provides an intuitive confidence measure for inferred clades \cite{Yang1997, Ronquist2012}. Popular software implementations include BEAST and MrBayes \cite{Drummond2012, Ronquist2012}, which both rely on variations of the Markov Chain Monte Carlo (MCMC) approach. Convergence issues have prompted the development of several other approaches, based, for example, on Sequential Monte Carlo \cite{Bouchard2012} and Stochastic Approximation Monte Carlo \cite{Cheon2008}.

Software like MEGA and PAUP* \cite{Tamura2013, Swofford2003} have made maximum likelihood (ML) phylogenetic reconstruction a popular alternative. RAxML \cite{Stamatakis2014} and FastTree \cite{Price2010} are more recent options, designed specifically to handle large datasets. They both rely on heuristic tree-searching strategies to considerably speed up likelihood optimization. Generally, methods for maximum likelihood phylogenetic reconstruction do not yield measures of confidence for clades, which are necessary to apply conventional clustering rules. To solve that problem, they are combined with a bootstrap scheme. However, the interpretation of bootstrap support for clades remains controversial \cite{Erixon2003, Susko2009, Makarenkov2010}.

Bayesian and ML phylogenetic approaches involve generating a large collection of trees. The maximum posterior probability (MAP) or ML estimate are natural choices for the tree that best describes the ancestry of the data. However, especially in large samples, the score for those estimates may not be much higher than that for many other trees. Therefore, summarizing a collection of phylogenies by building a so-called \textit{consensus tree} \cite{Larget1999, Bryant2003, Holder2008} is common. Unlike conventional point estimates, consensus trees provide measures of uncertainty for elements in the tree \textit{topology}, an unambiguous representation of the hierarchical nesting of clades in the phylogeny.

After computing a sensible phylogenetic estimate, one can then proceed to estimate clusters. \cite{Brenner2007} define a cluster as a clade known with high confidence, and with \textit{patristic distances} bounded above by a reasonably low value, where the patristic distance between any two sequences is calculated by summing branch lengths along the path linking the corresponding tips in the tree. The method itself however does not specify how confidence and distance requirements should be selected. In their ML-bootstrap analysis for example, \cite{Brenner2007} used a confidence threshold of $98$\% and a patristic distance requirement of $1.5$\%.

\cite{Prosperi2011} designed PhyloPart, a method that also defines clusters as clades known with high confidence. The genetic distance requirement is now formulated in terms of the median patristic distance in a clade. To conclude in clustering, we must have median patristic distance in a clade below a value equal to a reasonably low percentile of patristic distances in the entire tree. In their analyses, \cite{Prosperi2011} used the $1$st, $10$th, $15$th, and $30$th percentiles. The choice of a percentile threshold is arbitrary: in their study, it was selected to maximize agreement with a number of confirmed clusters.

Alternatively, \cite{Ragonnet-Cronin2013} proposed ClusterPicker, that also finds clusters by identifying clades inferred with reasonably high confidence. The distance requirement in ClusterPicker does not involve patristic distances, but rather simple pairwise estimates of genetic distance, computed for example with the JC69, K80, HKY85, or raw (Hamming) model \cite{Jukes1969, Kimura1980, Hasegawa1985}. The method is convenient, as it can be applied readily to consensus trees, which do not naturally have branch lengths. Once again, the tuning of the clustering requirements is left entirely to the investigator.

Clustering criteria are often arbitrary, and tend to be poorly justified. In Bayesian phylogenetic clustering, posterior probability requirements of $1$ are the most common \cite{Yang2006, Erixon2003}, although studies may opt for a lower value \cite{Ahumada-Ruiz2009}. In the ML-bootstrap framework, clade support requirements as low as $70$\% \cite{Chaix2003, Ibe2008, Lindstroem2006}, or above $90$\% \cite{LeighBrown2011, Ragonnet-Cronin2013, Brenner2007} are common. A lot of variability is also observed in genetic distance requirements. For instance, \cite{LeighBrown2011} use the HKY+$\gamma$ model \cite{Hasegawa1985} to assess pairwise distances between sequences and impose a maximum distance of $4.5$\% within any cluster. \cite{Prosperi2011} instead find that a median patristic distance requirement of $7$\% maximizes correspondence with known clusters.

The variety of standards encountered in the literature may reflect a lack of agreement as to what clusters correspond to \cite{Chalmet2010}. More recently, \cite{Vrbik2015} proposed the \textit{Gap Procedure}, a distance-based clustering approach that avoids phylogenetic reconstruction and cutpoint selection altogether by defining clusters based on a measure of distinctiveness. Although it is very fast, it does not provide any means to evaluate uncertainty around its point estimates. Like the Gap Procedure, the method presented in this paper aims to avoid cutpoint selection by giving clusters a straightforward definition. However, it should also offer an intuitive measure of confidence in cluster estimates. We designed it specifically for clustering HIV-1 sequencing data, which will be the main substantive focus in the remainder of the paper. 

\section*{Methods}

\begin{figure}[h]
  \centering
  \includegraphics[width = 7cm]{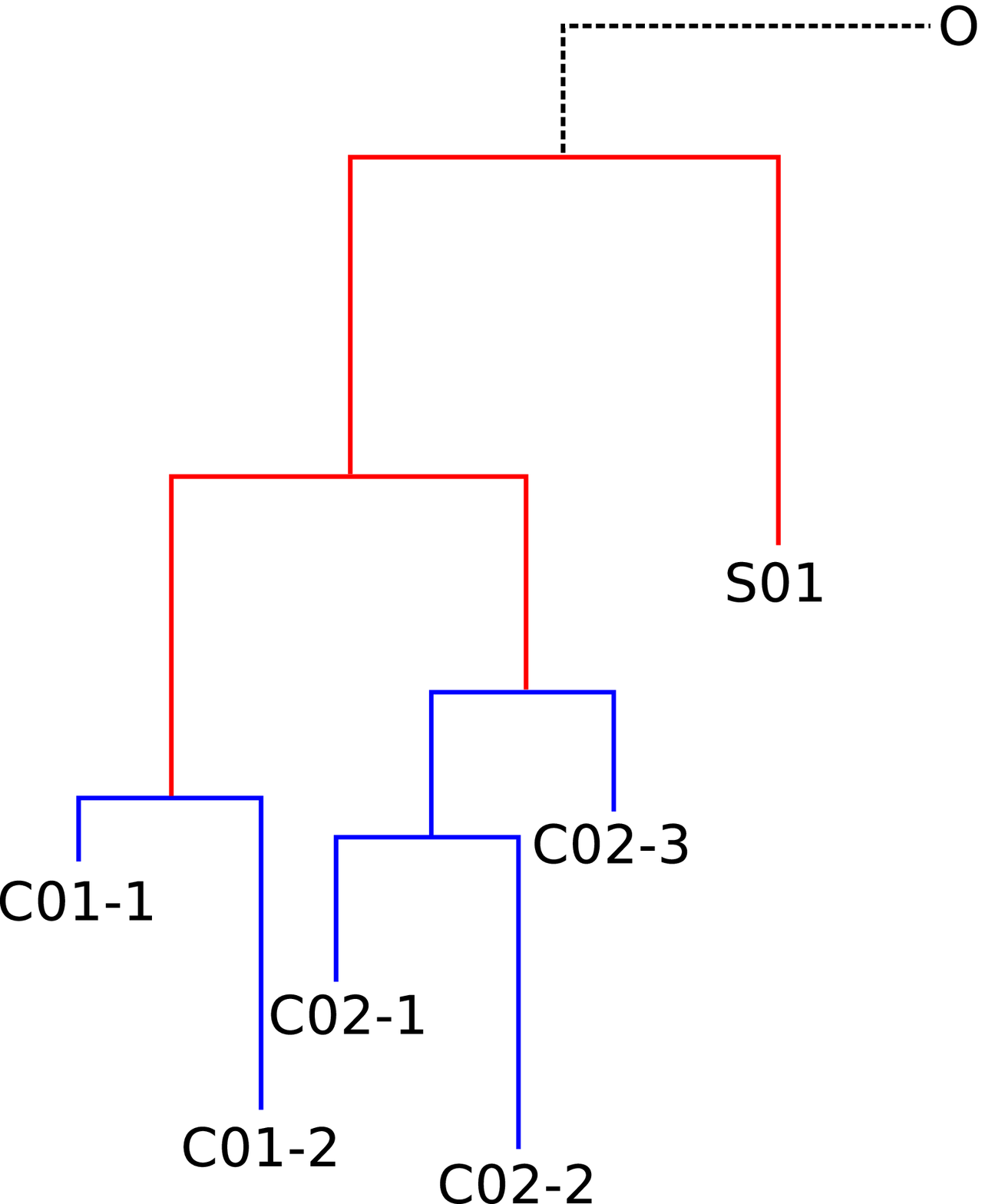}
  \caption[A phylogeny split into between- and within- cluster components]{\textbf{A phylogeny split into between- and within- cluster components}. Sequences C01-1 and C01-2 belong to cluster $1$, while C02-1, C02-2, and C02-3 belong to cluster $2$. Sequence S01 is a singleton, that is, a cluster of size $1$, and O is an outgroup, used to root the sample phylogeny. The red sub-phylogeny is called the \textit{between-cluster} phylogeny, while the blue sub-phylogenies are called the \textit{within-cluster} phylogenies.} 
  \label{fig:phyloComponents}
\end{figure}

DM-PhyClus is a MCMC-based algorithm \cite{Hastings1970} that innovates by relying on a definition of transmission clusters that better reflects clinical understanding, and by avoiding ad hoc distance and confidence requirements. DM-PhyClus makes use of a likelihood formulation that distinguishes between \textit{between-cluster} and \textit{within-cluster} components of the phylogeny, cf. Figure \ref{fig:phyloComponents}. The between-cluster phylogeny represents the ancestral relationships between each cluster's most recent common ancestor (MRCA), and the within-cluster phylogenies, the ancestral history of each cluster. 

Under DM-PhyClus, clusters have a clear definition: they are sets of sequences whose ancestral history is characterized by a specific distribution for branch lengths. In order for clusters to reflect quick transmission chains, we attribute branch lengths in the within-cluster phylogenies a prior with a reasonably low mean, in comparison to that for branches in the between-cluster phylogeny.

\subsection*{Likelihood}

We compute the tree likelihood recursively with Felsenstein's tree-pruning algorithm \cite{Felsenstein1981}. Let $(y_1, \dots, y_n)$ denote the sequence data, and $y_{i,s}$, the state at the $s$'th site, $s = 1, \dots, S$, of sequence $i$. If sequences are made up of nucleotides, $y_{i,s}$ can take one of $4$ values, each represented by a unit vector of length $4$. For example, nucleotides $A$ and $T$ are represented by vectors $(1,0,0,0)$ and $(0,1,0,0)$, respectively. 

At each site, evolution along branches of the tree, whose topology and branch lengths are denoted $\tau$ and $\bm{l}$, respectively, follows a continuous time Markov chain with rate matrix $Q$. Further, we assume that among-loci variation in evolution rates follows a discrete gamma distribution with $n_r$ categories and parameter $r$. Evolution occurs independently at different loci and so, the likelihood takes value,
\begin{equation}
 \zeta(\tau, \bm{l}, n_r, r, Q) = \prod_{s = 1}^S \zeta_s(\tau, \bm{l}, n_r, r, Q), \label{eq:basicLik}
\end{equation}
where $\zeta_s(\tau, \bm{l}, n_r, r, Q)$ represents the likelihood contribution of site $s$.

Let $j$ and $k$ index the two children of an arbitrary internal node $i$ in topology $\tau$, and $x_.$ be a numerical code for the state at node $.$, e.g. $A = 1$, $C = 2$, $T = 3$, $G = 4$. We have,
\begin{equation}
  L_{x_i}^{(s,i,m)} = \sum_{x_j} p_{x_i,x_j}(\xi_m l_j)L_{x_j}^{(s,j,m)} \sum_{x_k} p_{x_i, x_k}(\xi_m l_k)L_{x_k}^{(s,k,m)}, \label{eq:completeLik}
\end{equation}
where $p_{x_i,x_.}(\xi_m l_.)$ represents the transition probability from state $x_i$ to $x_.$ along a branch of length $l_.$, with coefficient $\xi_m$ being a scaling factor resulting from the conditioning on rate variation category $m$. We note that $x_i$ indexes the $\bm{L}^{(s,i,m)}$ vector, and it follows that the vector has as many elements as there are states in the data, e.g. $4$ for nucleotide data. From the Markov assumption, it follows that,
\begin{equation*}
 p_{x_i,x_.}(\xi_m l_.) = \exp(Q\xi_m l_.).
\end{equation*}
When index $i$ is for a tip, we have that $\bm{L}^{(s,i,m)} = y_{i,s}$. We must compute $\bm{L}^{(s,i,m)}$ for each combination of locus $s$, node $i$, and rate variation category $m$.

We start by computing $\bm{L}^{(s,i,m)}$ for all nodes $i$ whose children $j$ and $k$ are both tips. Then, we list all pairs of nodes $j$ and $k$ for which both $\bm{L}^{(s,j,m)}$ and $\bm{L}^{(s,k,m)}$ are known, and compute $\bm{L}^{(s,i,m)}$ for each of them.

Let the root of the tree have index $\vartheta$. We have that the likelihood contribution of site $s$ takes value,
\begin{equation*}
  \zeta_s(\tau, \bm{l}, n_r, r, Q) = \dfrac{1}{n_r}\sum_{m = 1}^{n_r} \sum_{x_{\vartheta}} L_{x_\vartheta}^{(s,\vartheta,m)}p_{x_{\vartheta}},
\end{equation*}
where $\bm{p}$ represents the limiting probabilities of the Markov chain.

In real DNA sequences, sequencing may reveal that two or more nucleotides can be found at certain loci, producing an \textit{ambiguity}. In Felsenstein's tree-pruning algorithm, ambiguities are expressed as a sum of the unit vectors for the potential states. For example, if A and T are observed at site $m$ in sequence $i$, we get that $y_{i,m} = [1,1,0,0]$.       

\subsection*{Priors}

We denote branch lengths in the within-cluster and between-cluster components $\bm{l}^{(w)}$ and $\bm{l}^{(b)}$, respectively. We assign branch lengths in the between-cluster phylogeny a log-normal prior with parameters $\mu$ and $\sigma$. We picked that distribution because of its potentially heavy right tail, which allows for a small number of distinctively long branches. We tune priors for those parameters based on a desired mean and coefficient of variation. To lighten the computational load, we assign that mean a uniform prior over a finite number of discrete values, and the coefficient of variation is fixed. We assign branch lengths in within-cluster phylogenies an exponential prior with rate $\delta$, whose prior is, like before, discrete uniform over a finite range of sensible values.

We assign cluster membership indices $(c_1, \dots, c_n)$ a multinomial prior with probability parameters $(p_1, \dots, p_{\max(\bm{c})})$, weighted by values from a Poisson distribution, with rate parameter $\lambda$, evaluated at $\max(\bm{c})$ and an indicator function giving probability $0$ to configurations not meeting the clade assumption,
\begin{equation}
  \begin{aligned}
  P(c_1, \dots, c_n \mid \tau, \lambda, \bm{\pi}) &\propto \binom{n}{n_1 \dots n_{\max(\bm{c})}} \pi_1^{n_1} \cdots \pi_{\max(\bm{c})}^{n_{\max(\bm{c})}} \dfrac{\exp(-\lambda)\lambda^{\max(\bm{c})}}{\max(\bm{c})!} \times \\ 
  &\times I[\mbox{Partition allowed by $\tau$}],
  \end{aligned} \label{eq:condVectorC}
\end{equation}
with $n_k = \sum_{i=1}^n I[c_i = k]$ and $I[.]$ being an indicator function.

\begin{figure}[h]
  \centering
  \includegraphics[width = 10cm]{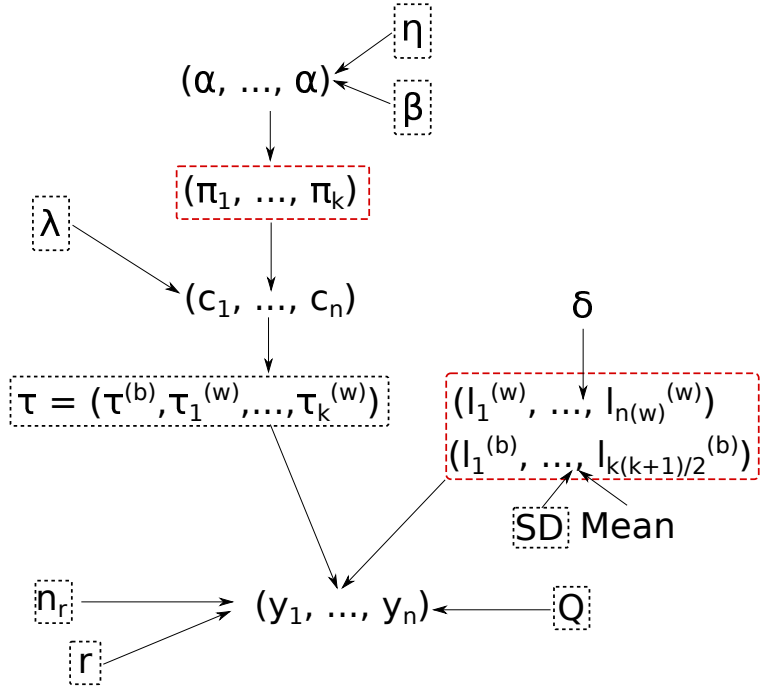}
  \caption[Graphical representation of the relationships between parameters and the data]{\textbf{Graphical representation of the relationships between parameters and the data}. Parameters in a black box are fixed. Parameters in a red box are marginalized out. The vector $(y_1, \dots, y_n)$ is the sample, and ``SD'' stands for standard deviation. We denote the within-cluster phylogenies $(\tau^{(w)}_1, \dots, \tau^{(w)}_k)$, $k$ being the number of clusters, and the between-cluster phylogeny, $\tau^{(b)}$. Within-cluster phylogenies are degenerate when they support a cluster of size $1$, while the between-cluster phylogeny is degenerate when the sample comprises only one cluster. The log-normal prior distribution for the between-cluster branch lengths is reparameterized in such a way that it has mean and standard deviation parameters, like in the normal distribution.} 
  \label{fig:parameters}
\end{figure}

The probability parameters have a symmetric Dirichlet hyperprior with concentration parameter $\alpha$, to which we assign a gamma hyperprior with shape and scale parameters $\eta$ and $\beta$. We summarize parameters in Figure \ref{fig:parameters}.

\subsection*{Posterior probability derivation}

We are interested primarily in the posterior distribution of cluster membership indices $\bm{c}$ and so, we marginalize out probability parameters $\bm{\pi}$, as well as all branch lengths. Marginalizing out $\bm{\pi}$ from Equation \ref{eq:condVectorC}, we obtain,
\begin{equation}
  \begin{aligned}
    P(c_1, \dots, c_n \mid \tau, \bm{\alpha}, \lambda) &\propto \dfrac{B(\bm{n} + \bm{\alpha})}{B(\bm{\alpha})}\binom{n}{n_1 \dots n_{\max(\bm{c})}} \dfrac{\lambda^{\max(\bm{c})} \exp(-\lambda)}{\max(\bm{c})!} \times \\ 
    &\times I[\mbox{Partition allowed by $\tau$}], \label{eq:shortCondPrior}
  \end{aligned}
\end{equation}
with,
\begin{equation*}
  B(\bm{\alpha}) = \dfrac{\prod_{i=1}^{\max(\bm{c})} \Gamma(\alpha_i)}{\Gamma\left(\sum_{i=1}^{\max(\bm{c})} \alpha_i\right)}.
\end{equation*}

We use Monte Carlo integration to marginalize out branch lengths from the likelihood. When the number of Monte Carlo replications $K$ is large enough, the probability of a transition from state $x_i$ to $x_j$ over any given branch is approximately,
\begin{equation}
  P(x_j \mid x_i, \bm{c}) = \int_{\mathcal{D}(l\mid\bm{c})} [\exp(Ql)]_{(x_i,x_j)} p(l \mid \bm{c}) dl \approx \dfrac{1}{K}\sum_{k = 1}^K [\exp(Ql_k)]_{(x_i,x_j)}, \label{eq:margTransProb}
\end{equation}
where $\mathcal{D}(l\mid\bm{c})$ is the domain of $l \mid \bm{c}$, $p(l \mid \bm{c})$ is the prior distribution of $l$ conditional on $\bm{c}$, and $l_k$ is drawn from that distribution. $[\exp(Ql)]$ denotes the transition probability matrix along a branch of length $l$, and $[\exp(Ql)]_{(x_i,x_j)}$ represents element $(x_i,x_j)$ of that matrix. The conditioning on $\bm{c}$ appears as a result of the marginalization, because of the different priors for branch lengths in the within-cluster phylogenies and the between-cluster phylogeny.

The posterior distribution of the cluster membership indices is denoted,
\begin{equation*}
  P(c_1, \dots, c_n \mid y_1, \dots, y_n, \tau, \bm{\alpha}, \lambda) \propto \zeta(\tau, n_r, r, Q \mid c_1, \dots, c_n)  P(c_1, \dots, c_n \mid \tau, \bm{\alpha}, \lambda),
\end{equation*}
where $P(c_1, \dots, c_n \mid \tau, \bm{\alpha}, \lambda)$ is given by Equation \ref{eq:shortCondPrior} and $\zeta(\tau, n_r, r, Q \mid c_1, \dots, c_n)$ is obtained by replacing $p_{x_i,x_.}(\xi_m l_.)$ in Equation \ref{eq:completeLik} by the approximation derived in Equation \ref{eq:margTransProb}, but with simulated branch lengths $l_k$ being multiplied by $\xi_m$. There is a one-to-one correspondence between $(c_1, \dots, c_n)$ and the breakdown of $\tau$ into within-cluster phylogenies and between-cluster phylogeny, and the conditioning on $(c_1, \dots, c_n)$ in the marginal likelihood appears as a result.

\subsection*{Transition kernels and Metropolis-Hastings (MH) ratios}

DM-PhyClus first searches for a sensible phylogenetic estimate, that acts to restrict the space of potential cluster membership indices, and then, conditional on that phylogeny, performs successive Metropolis-Hastings (MH) updates of the concentration parameter and the cluster membership indices.

We sample tentative transitions in the space of concentration parameter $\alpha$ from a uniform distribution defined over an interval of length $1$ centered around the current value of $\alpha$, resulting in the transition kernel ratio reducing to $1$. We propose moves in the space of cluster membership indices $\bm{c}$ by using a cluster split-merge strategy. Any cluster of size $2$ or more can be split in two disjoint clusters, corresponding to the clades supported by the children of the original cluster's root. We can merge any two neighbouring clusters, or in other words, any two clusters whose most recent common ancestor is at most one split above their respective roots. The transition kernel is a discrete uniform distribution over all split-merge transitions allowed by the topology from the current state. It follows that the transition kernel ratio is equal to the total number of potential moves from the current configuration divided by the total number of potential moves starting from the proposal. With the ratio of priors obtained from Equation \ref{eq:shortCondPrior} and the conventional likelihood ratio, we have all necessary components for computing the MH ratio.

\subsection*{Point estimates for cluster membership indices}

We produce two kinds of estimates for cluster membership indices, the \textit{maximum posterior probability} (MAP) estimate, and the \textit{linkage-xx} estimate, which we obtain in three steps,
\begin{enumerate}
  \item Derive an \textit{adjacency matrix} from each sampled cluster membership indices vector.
  \item[] An adjacency matrix is a symmetrical matrix with a $1$ at position $(i,j)$ if elements $i$ and $j$ co-cluster, and with a $0$ otherwise. \label{enum:getAdjMat} 
  \item Average adjacency matrices computed in step \ref{enum:getAdjMat} and apply a co-clustering frequency threshold of $xx$\%. \label{enum:meanAdjMat}
  \item[] The average adjacency matrix provides co-clustering frequencies. All frequencies higher than the threshold are rounded up to $1$, while all others are rounded down to $0$. 
  \item Identify all \textit{disjoint} sets, called \textit{modules} or \textit{components}, from the matrix obtained in step \ref{enum:meanAdjMat}.
  \item[] Two sets of sequences are disjoint if no co-clustering exists between them. We use the walktrap algorithm \cite{Pons2005} to detect disjoint sets, which leads to the cluster estimates. 
\end{enumerate}

We present a structured, step-by-step description of DM-PhyClus in Supplementary Material S1.

\subsection*{Simulation study}

\subsubsection*{Data}

We simulate an HIV-1 sequence dataset of size $200$ by going through the following steps:
\begin{enumerate}
 \item Sample the total number of clusters from a Poisson distribution with mean $50$,
 \item Sample cluster assignment probabilities from a symmetric Dirichlet distribution with a concentration parameter generated from a normal distribution with mean $10$ and standard deviation $2$,
 \item Sample $200$ values from a multinomial distribution with the obtained probability vector,
 \item Generate each within-cluster phylogeny by picking a topology at random, and by sampling branch lengths from an exponential distribution with mean equal to $0.003$,
 \item Generate the between-cluster phylogeny by picking a topology at random, and by sampling branch lengths from a log-normal distribution with mean and standard deviation equal to $0.008$,
 \item Let the \textit{HXB2} sequence evolve along the simulated tree, with evolution rate matrix and limiting probabilities obtained from \cite{Posada2001}.
\end{enumerate}
HXB2 is an HIV-1 subtype B sequence that serves as a reference for site position numbers in any HIV-1 sequence. In other words, the range of site indices in any HIV-1 sequence is found by aligning it with HXB2. We generate $50$ datasets in total, and add to each of them an arbitrary subtype C outgroup (\url{http://www.hiv.lanl.gov/}, accession number: AB254141) for rooting the inferred phylogenies. We list parameters used for data generation in Supplementary Material S2.

\subsubsection*{Scenarios}

Assessing sensitivity of the cluster estimates to the concentration parameter prior is vital, as it may be challenging to properly specify in practice. For each simulated dataset, we run DM-PhyClus under the assumption that the concentration parameter follows a gamma distribution with scale parameter $0.1$, and, successively, with means $1$, $10$, and $100$. The use of fixed estimates for the mutation rate matrix and limiting probabilities may also affect cluster recovery. To verify that such a restriction is not overly detrimental to cluster recovery, we use values for those parameters obtained from a separate analysis of a real HIV-1 sequence dataset, that we ensure are reasonably different from those used for data generation.

\subsubsection*{Setup}

Given the synthetic nature of the problem, tuning priors for branch lengths is difficult and so, we opt for an empirical Bayes approach, where we use maximum likelihood phylogenetic estimates to derive mean branch lengths in the within- and between-cluster phylogenies. We then define a range around each of the obtained means with radius equal to $8\%$ of the obtained mean. Finally, we select $20$ equidistant points in each range, at which we compute transition probability matrices by sampling $100,000$ values from the log-normal distribution for between-cluster branch lengths, or the exponential distribution for within-cluster branch lengths.

We use RAxML \cite{Stamatakis2014} to obtain an estimate of the maximum likelihood phylogeny, and to perform $500$ bootstrap iterations, producing the usual clade support estimates. We then get starting values for the cluster membership indices by running a depth-first search on the tree. We stop exploration along any path once we find a clade with bootstrap support greater than $70$\% and with patristic distances below a certain threshold, selected through maximization of the Dunn index \cite{Dunn1973}, a measure of clustering quality. In a first round of simulations, we use that partition as a starting value for the chain, and the maximum likelihood topology to bound the space of cluster solutions.

In a second round of simulations, before launching the main chain, we explore the topological space around the maximum likelihood phylogeny, using nearest-neighbour interchange to find a configuration that improves posterior probability, and letting values for the concentration parameter and cluster membership indices vary as well. We start the MCMC run once a suitable topology is identified. We present an exhaustive list of the tuning parameter values used in the simulations in Supplementary Material S2.

\subsubsection*{Chain configuration and point estimates comparison}

For each simulated dataset, we produce $55,000$ samples from the posterior distribution of the cluster membership indices vector. We apply a thinning ratio of $1$ over $50$, and take out the first $5,000$ iterations as a burn in, leaving us with $1000$ samples. Once the MCMC run is complete, we obtain the MAP and linkage-$xx$ cluster estimates, and measure overlap between the real and inferred clusters with the adjusted Rand Index (ARI), a measure of similarity between two sets of clusters. It involves the ratio of pairs of elements that are similarly co-clustered or dissociated in both sets to the total number of pairs in the sample, combined with a numerical adjustment for chance. It is bounded above by $1$, which indicates perfect correspondence. We compare those estimates to those we initially obtained from RAxML, which we refer to as the \textit{Bootstrap-70} estimates, and to the estimates from the so-called \textit{Gap procedure}, a quick distance-based genetic sequence clustering approach that requires minimal tuning \cite{Vrbik2015}. The Bootstrap-70 estimate is a natural standard for comparison, since it is obtained by applying a conventional method for the clustering of HIV-1 sequencing data \cite{Erixon2003}.     

\subsection*{Real data analysis}

\subsubsection*{Data}

The original sample consists of $3,537$ HIV-1 subtype B sequences collected for the Qu\'{e}bec HIV genotyping program \cite{Brenner2007}. Each sequence is from a different male patient belonging to the injection drug user (IDU) or men who have sex with men (MSM) risk category, and that has not yet started antiretroviral therapy, the standard treatment regimen for HIV-positive individuals. The dataset includes sites $10$-$297$ of the protease region (PR), and $112$-$741$ of the reverse transcriptase (RT) region, of the \textit{pol} gene.

\cite{Brenner2011a} obtained an initial set of clusters by partitioning the sample through inspection of the maximum likelihood tree, selecting clades with bootstrap support greater than $98$\% and whose patristic distances were below $1$\%. They also looked for congruent polymorphisms and mutational motifs. Whenever new sequences entered the database, they updated their cluster estimates by re-inferring the tree, and attaching new sequences to previously-inferred clusters when the clade they belonged to had bootstrap support greater than $98$\%. They also used clinical and demographic information to exclude sequences from inferred clusters.

We focus on a subsample of $526$ sequences, made up of $18$ previously-inferred clusters of sizes ranging from $2$ to $69$, inclusively, as well as $12$ singletons selected uniformly at random in the original sample. We add to the sample $3$ subtype C outgroups from Zambia, downloaded from the Los Alamos HIV-1 sequencing database (\url{http://www.hiv.lanl.gov/}, accession numbers AB254141, AB254142, AB254143).

\subsubsection*{Bootstrap analysis}

To evaluate sensitivity of DM-PhyClus to the input topology, we produce $100$ bootstrap samples of the data by resampling site indices with replacement and re-assembling each sequence based on the sampled indices. We use maximum likelihood topological estimates and use the same strategy as in the simulations to obtain starting values for the chain. Each run also consists of $55,000$ iterations, with a burn-in of $5,000$ and a thinning ratio of $1/50$.

\subsubsection*{Approximation of the fully Bayesian analysis}

Fixing the topological parameter in the chain results in the inference not being fully Bayesian. Such an approximation is acceptable only so long as we can establish that the results do not differ too much from those resulting from the fully Bayesian approach. To do so, we first use MrBayes \cite{Ronquist2012}, run under the default configuration, to sample $1.5$ million phylogenies from the posterior distribution $P(\tau \mid \bm{y}, \dots)$, where $\dots$ represents the other parameters. We take out the first $375,000$ samples as a burn-in, and apply a thinning ratio of $1/500$. Of the remaining $2,250$ samples, we select $100$ uniformly at random, which we use as input in $100$ separate runs of DM-PhyClus. Each run produces samples from the conditional posterior distribution of the cluster membership indices $P(\bm{c} \mid \tau_i, \dots, \bm{y})$, $i = 1, \dots, 100$. Noting that,
\begin{equation*}
 P(\bm{c} \mid \bm{y}) = E_{\tau}[P(\bm{c} \mid \tau, \bm{y})] \approx \sum_{i = 1}^{100} P(\bm{c} \mid \tau_i, \bm{y})/100,
\end{equation*}
we see that high overlap between the maximum posterior probability cluster membership indices obtained from the $100$ chains ensures that the peak of $P(\bm{c} \mid \bm{y})$ is found at a configuration similar to those obtained in each individual run, thus confirming the quality of the approximation resulting from the conditioning assumption.

\subsubsection*{Main run}

We obtain starting values with the help of RAxML, under the assumption that genetic distance follows the GTR + $\Gamma(3)$ model. As in the simulations, we configure priors for branch lengths based on the maximum likelihood topology. We use limiting probabilities and nucleotide substitution rates previously inferred for HIV-1 subtype B \cite{Posada2001}. We assume discrete gamma substitution rate variation with $3$ categories. Finally, we fix the rate parameter for the Poisson distribution at $30$, the number of clusters obtained in \cite{Brenner2011a}. We run $220,000$ iterations, keeping one iteration out of $150$ and taking out the first $70,000$ iterations as a burn-in. We then obtain point estimates for cluster membership indices as before. An exhaustive list of tuning parameter values used in all real data analyses is available in Supplementary Material S3.

\subsection*{Software}

We present a technical description of the software in Supplementary Material S4. We implement the algorithm in R, with functions contained in the \textit{phangorn}, \textit{ape}, and \textit{phytools} libraries \cite{Schliep2011, Revell2012}. Likelihood evaluations rely on compiled C++ code integrated into the R script using the \textit{Rcpp} and \textit{RcppArmadillo} packages \cite{Eddelbuettel2011, Eddelbuettel2013}. We produce starting values with RAxML \cite{Stamatakis2014}. Finally, we also produce cluster estimates with the the GapProcedure package \cite{Vrbik2015}. A package, \textit{DMphyClus}, is available on Github (\url{https://github.com/villandre/DMphyClus}) and will be submitted to CRAN.

\section*{Results}

\subsection*{Simulation study} % Add table with precise parameters used in the simulations.

\begin{table*}
  \centering
    \caption[Summary statistics for adjusted Rand indices (ARI) for cluster membership estimates obtained from chains run on 50 datasets under different simulation scenarios.]{\textbf{Summary statistics for adjusted Rand indices (ARI) for cluster membership estimates obtained from chains run on $50$ datasets under different simulation scenarios.}}
    \resizebox{\textwidth}{!}{    
    \begin{tabular}{ccccccccccc}
      \hline \hline      
      \textbf{Topology used} & \textbf{Alpha mean} & \textbf{Estimator} & \textbf{Min.} & \textbf{Max.} & $\bm{10}$\textbf{th perc}. & \textbf{Median} & $\bm{90}$\textbf{th perc.} & \textbf{Mean} & \textbf{SD} & \textbf{SE} \\ \hline \hline  
      GapProcedure& - & - & 0.012 & 0.719 & 0.030 & 0.385 & 0.654 & 0.361 & 0.227 & 0.005 \\ \hline
      Bootstrap-70& - & - & 0.074 & 0.882 & 0.256 & 0.483 & 0.771 & 0.504 & 0.221 & 0.004 \\ \hline      
      ML topology& 10 & MAP & 0.000 & 0.935 & 0.686 & 0.820 & 0.900 & 0.769 & 0.210 & 0.004 \\
      && Linkage-0.7 & 0.000 & 0.946 & 0.711 & 0.853 & 0.920 & 0.793 & 0.213 & 0.004 \\
      && Linkage-0.8 & 0.000 & 0.971 & 0.707 & 0.838 & 0.912 & 0.793 & 0.213 & 0.004 \\
      && Linkage-0.9 & 0.000 & 0.962 & 0.710 & 0.822 & 0.893 & 0.771 & 0.206 & 0.004 \\
      && Linkage-1 & 0.089 & 0.710 & 0.359 & 0.494 & 0.631 & 0.484 & 0.129 & 0.003 \\
      \cline{2-11}
      & 1 & MAP & 0.098 & 0.862 & 0.328 & 0.619 & 0.833 & 0.601 & 0.199 & 0.004 \\
      && Linkage-0.7 & 0.012 & 0.939 & 0.381 & 0.725 & 0.861 & 0.653 & 0.218 & 0.004 \\
      && Linkage-0.8 & 0.011 & 0.959 & 0.394 & 0.760 & 0.865 & 0.680 & 0.207 & 0.004 \\
      && Linkage-0.9 & 0.053 & 0.937 & 0.466 & 0.776 & 0.885 & 0.712 & 0.191 & 0.004 \\
      && Linkage-1 & 0.159 & 0.716 & 0.397 & 0.470 & 0.646 & 0.491 & 0.103 & 0.002 \\ \cline{2-11}
      & 100 & MAP & 0.123 & 0.931 & 0.594 & 0.848 & 0.917 & 0.790 & 0.196 & 0.004 \\ 
      && Linkage-0.7 & 0.123 & 0.973 & 0.346 & 0.859 & 0.931 & 0.791 & 0.215 & 0.004 \\
      && Linkage-0.8 & 0.123 & 0.971 & 0.348 & 0.852 & 0.920 & 0.785 & 0.211 & 0.004 \\
      && Linkage-0.9 & 0.123 & 0.980 & 0.378 & 0.820 & 0.896 & 0.761 & 0.202 & 0.004 \\
      && Linkage-1 & 0.123 & 0.802 & 0.351 & 0.514 & 0.652 & 0.504 & 0.133 & 0.003 \\
      \hline
      MAP topology& 10 & MAP & 0.000 & 0.935 & 0.714 & 0.839 & 0.923 & 0.798 & 0.180 & 0.004 \\ 
      && Linkage-0.7 & 0.000 & 0.950 & 0.727 & 0.858 & 0.919 & 0.818 & 0.172 & 0.004 \\
      && Linkage-0.8 & 0.000 & 0.953 & 0.791 & 0.846 & 0.919 & 0.823 & 0.165 & 0.003 \\
      && Linkage-0.9 & 0.000 & 0.947 & 0.751 & 0.824 & 0.891 & 0.798 & 0.156 & 0.003 \\
      && Linkage-1 & 0.000 & 0.686 & 0.318 & 0.449 & 0.598 & 0.454 & 0.117 & 0.002 \\
      \cline{2-11}
      & 1 & MAP & 0.011 & 0.870 & 0.329 & 0.623 & 0.832 & 0.598 & 0.203 & 0.004 \\
      && Linkage-0.7 & 0.162 & 0.930 & 0.321 & 0.738 & 0.848 & 0.649 & 0.212 & 0.004 \\
      && Linkage-0.8 & 0.170 & 0.931 & 0.384 & 0.746 & 0.872 & 0.671 & 0.201 & 0.004 \\
      && Linkage-0.9 & 0.175 & 0.911 & 0.437 & 0.764 & 0.852 & 0.693 & 0.178 & 0.004 \\
      && Linkage-1 & 0.341 & 0.745 & 0.396 & 0.516 & 0.660 & 0.524 & 0.093 & 0.002 \\ \cline{2-11}
      & 100 & MAP & 0.123 & 0.947 & 0.761 & 0.854 & 0.914 & 0.816 & 0.171 & 0.003 \\
      && Linkage-0.7 & 0.123 & 0.976 & 0.793 & 0.867 & 0.923 & 0.830 & 0.170 & 0.003 \\
      && Linkage-0.8 & 0.123 & 0.970 & 0.789 & 0.857 & 0.914 & 0.825 & 0.169 & 0.003 \\
      && Linkage-0.9 & 0.123 & 0.965 & 0.703 & 0.819 & 0.901 & 0.789 & 0.164 & 0.003 \\
      && Linkage-1 & 0.123 & 0.672 & 0.298 & 0.459 & 0.619 & 0.457 & 0.122 & 0.002 \\       
      \hline \hline
    \end{tabular}
  }
  \label{tab:ARIsSimulations}
\end{table*}

On an Intel(R) Xeon(R) CPU E7-4820 v4 \@ 2.00GHz CPU, running $55,000$ iterations took on average a bit more than $2$ hours. Log-posterior probability graphs show no obvious issue with autocorrelation or convergence, and indicate good mixing (see, for example, Supplementary Material S5). We show the obtained ARIs for the six scenarios in table \ref{tab:ARIsSimulations}. Overall, mean cluster recovery from DM-PhyClus was superior than that from the conventional Bootstrap-70 approach and GapProcedure, both of which usually struggled to recover the clusters. We observe a noticeable drop in mean overlap when the concentration parameter has a prior whose mean is much smaller than that used for data generation, but not when it is larger.

The linkage-$xx$ estimates performed comparably or slightly better than the MAP estimates when the linkage requirement was $0.7-0.8$ and the prior on the concentration parameter had mean equal or superior to the the true value. When the prior underestimated the true concentration parameter value however, the linkage estimates greatly improved recovery, sometimes as much as much as $10$\%, as long as the linkage requirement was not $1$. Maximum observed recovery rates were also consistently superior for the linkage estimates.

The slightly better performance of DM-PhyClus when the concentration parameter has a mean greater than that used for data generation was unexpected. We observe it both when the MAP and ML topologies are used. When the concentration parameter prior had mean $10$, two chains returned a MAP configuration with a single cluster, producing the $0$ in the table, which explains at least part of the gap. The datasets analysed by those chains seem to imply a hard clustering problem, as evidenced by the low recovery rates from Bootstrap-70, $0.13$ and $0.18$. Overall, starting with the MAP configuration from a shorter preliminary run resulted in small increases in mean recovery rates. When the concentration prior mean was $10$, the same two chains as before resulted in a MAP configuration with only $1$ cluster, yielding ARI = $0$. With median recovery around $0.87$ in the better scenarios, we are not overly worried about the consequences of using fixed values for the limiting probabilities and mutation rate matrices, as long as they are selected reasonably.

\subsection*{Real data analysis}

\subsubsection*{Bootstrap analysis}

We measured overlap within all pairs of MAP configurations produced in the bootstrap analysis. ARIs ranged from $0.10$ to $0.98$, with median $0.83$ and mean $0.72$, indicating reasonable robustness of the chain to the assumed topology. Unsurprisingly, linkage estimates led to essentially the same conclusion. For example, overlap between cluster configurations proposed under the linkage-$70$ estimate ranged from $0.11$ to $0.98$, with median $0.83$ and mean $0.70$. Moreover, concordance between MAP estimates from the bootstrap replicas and the MAP cluster configuration obtained from the full data was generally high, with median and mean ARI equal to $0.88$ and $0.80$, respectively.

\subsubsection*{Approximation of the fully Bayesian analysis}

Estimates based on the $100$ topologies sampled with MrBayes were overall very similar, leading to the conclusion that the DM-PhyClus estimates are reasonable approximations of those resulting from a fully Bayesian analysis. Indeed, concordance between the MAP estimates obtained from the $100$ chains tended to be high: ARIs ranged from $0.38$ to $1$, with median and mean $0.89$ and $0.86$, respectively. Overlap with the usual MAP estimate, obtained conditional on the topology found to optimize joint posterior probability after a short exploration of the topological space, was also considerable, with median and mean $0.92$ and $0.90$, respectively.

\subsubsection*{Full data analysis}

\begin{figure}[ht]
  \centering
  \includegraphics[width = 10cm]{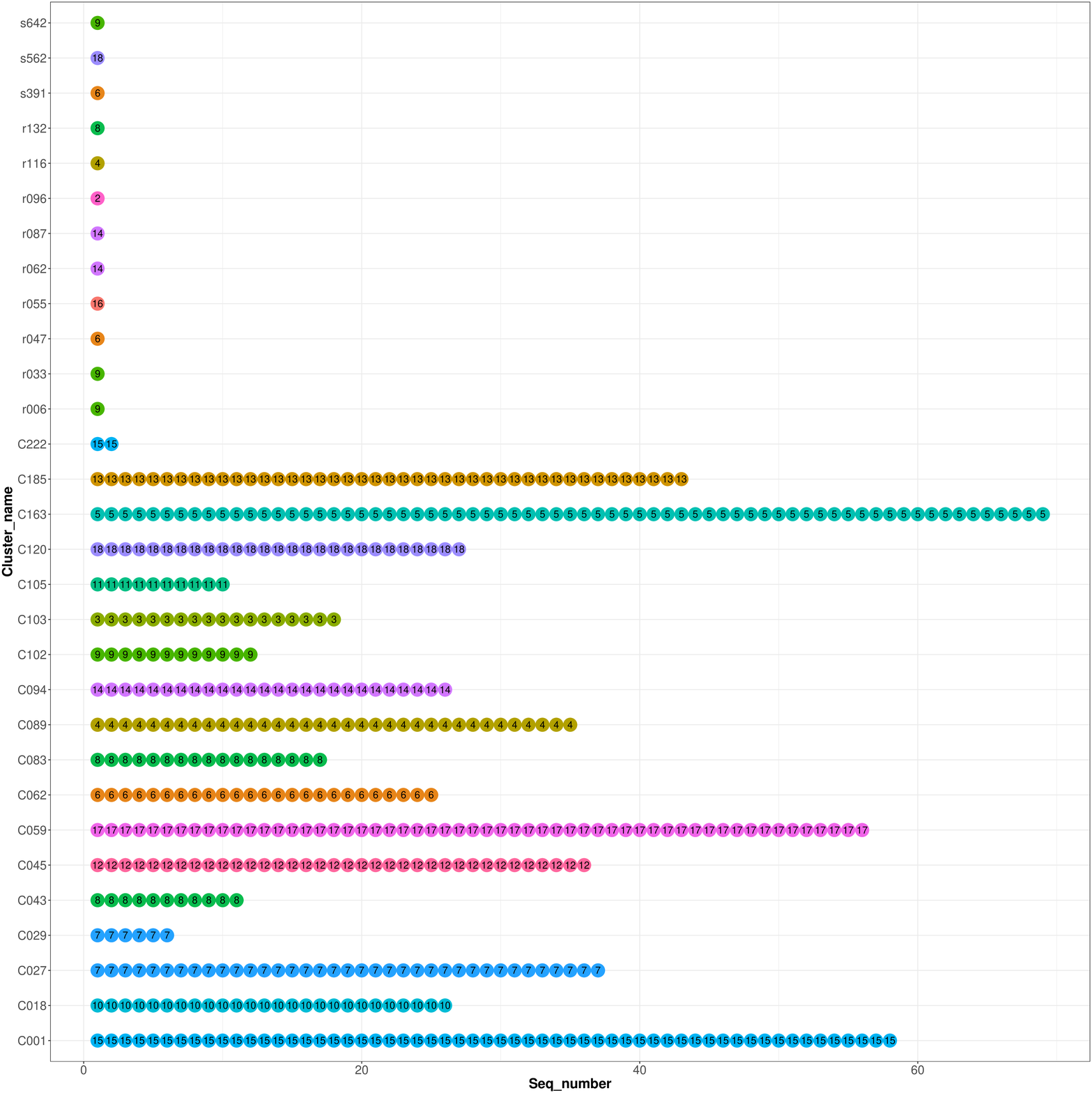}
  \caption[Comparison of the DM-PhyClus cluster estimates with a proposed cluster configuration for the real dataset.]{\textbf{Comparison of the DM-PhyClus cluster estimates with a proposed cluster configuration for the real dataset.} The coordinates on the vertical axis indicate cluster membership according to \cite{Brenner2011a}, and the colour and number of each dot, the cluster membership according to the maximum posterior probability (MAP) estimate of DM-PhyClus.}
  \label{fig:clusterVisualization}
\end{figure}

The MAP configuration obtained from DM-PhyClus revealed the existence of $16$ clusters of size $2$ or more, and $2$ singletons. Linkage estimates were identical to the MAP estimate when the linkage requirement was $98$\% or below, indicating little uncertainty in the returned partition. The Gap Procedure returned a rather similar set of clusters (ARI = $0.87$). We represent clusters from DM-PhyClus against those from the curated analysis in Figure \ref{fig:clusterVisualization}. DM-PhyClus has a tendency to merge neighbouring clusters, as evidenced by the smaller number of singletons and the merger of clusters $43$ and $83$, which also absorbed sequence r132, and of clusters $27$ and $49$. The GapProcedure, on the other hand, proposed a configuration with $43$ clusters of size $2$ or more, and $14$ singletons, splitting, for example, clusters $18$ and $59$ in $3$ and $8$ sets, respectively.

\section*{Discussion}

In this paper, we introduced a phylogenetic clustering algorithm, DM-PhyClus, that integrates an original cluster definition into cluster inference, which results in more intuitive estimates, unlike conventional approaches, that rely instead on arbitrary cutpoints applied a posteriori to a phylogenetic estimate. Simulations indicate that the algorithm can accurately recover phylogenetic clusters, often outperforming more conventional approaches. Analysis of a real dataset of HIV-1 subtype B sequences revealed a set of clusters largely similar to that from a previous analysis, but with more straightforward inference.

The study does have some limitations. Because of time constraints, we were only able to run short chains in the simulations. Log-posterior probability graphs for the simulated samples however did strongly suggest that the chains had converged, making us confident that increasing the number of iterations would not change our conclusions. We suspect that the apparent weakness of Bootstrap-70 might be in part attributable to the use of the Dunn index. For several simulated datasets, we noticed that it failed to identify the optimal partition in terms of recovery. Comparing our results to that solution would have been unfair, however, since identifying it requires knowledge of the true clusters. For computational reasons and to ensure adequate mixing in the chain, we opted for a fixed topology, thus limiting the number of partitions the algorithm can propose and ignoring uncertainty in phylogenetic reconstruction. Although simulations and the real data analyses indicate that this simplification works well in practice, proposing an efficient transition kernel that jointly updates cluster membership indices and the phylogeny would be necessary.

Further DM-PhyClus rests on the assumption that cluster-specific phylogenies have a distinctive branch length distribution. Our goal was to reflect intuitive understanding of transmission clusters, but our branch length assumptions do remain simplistic. Phylogenies for HIV-1, for instance, are characterized by long external branches \cite{Kouyos2011}. Moreover the exponential prior is known for producing overly long trees \cite{Wang2014}. The assumption however is common in Bayesian phylogenetic inference \cite{Drummond2012}, and leads to considerable computational simplifications. It is unclear whether more sophisticated, potentially dependent, branch length priors would improve cluster inference overall. Given the often high recovery rates observed in the simulations, we are confident that the simplification was not overly detrimental. Improvements to the code should also make it possible to apply DM-PhyClus to much larger datasets, such as those collected for major HIV-1 genotyping programs.

We contend DM-PhyClus is a worthwhile addition to existing methods used to detect transmission clusters. Understanding clustering in epidemics is crucial: in the case of HIV-1 among men who have sex with men for example, transmission clusters have been found to contribute overwhelmingly to incidence \cite{Brenner2011a, Brenner2013}. Investigations into the reasons behind the existence of those clusters are likely to help in reducing transmission rates, and those studies will need to rely on methods based on cluster definitions that reflect clinical insight, like DM-PhyClus.

% \section*{Ethics approval and consent to participate}
% 
% Ethics approval for the Quebec HIV genotyping program was obtained from individual study sites, the Laboratoire de sant\'{e} publique du Qu\'{e}bec, and the Quebec Ministry of Health committee on confidentiality and access of information.

% \section*{Author's contributions}
% 
% LV wrote the article, performed the simulations. LV, AL, and DAS jointly formulated the algorithm. AL and DAS suggested and reviewed analyses. BB and MR provided the HIV-1 sequences.

\section*{Data availability}

All simulated data generated or analysed during this study are included in this published article or on Zenodo (DOI 10.5281/zenodo.839849). The Qu\'{e}bec HIV genotyping program sequences cannot be made publicly available for confidentiality reasons. A small subset of sequences can be provided for verification purposes upon request. 

\section*{Acknowledgements}

This work was supported by a training award from the Fonds de recherche du Qu\'{e}bec-Sant\'{e} (FRQS), funding from the Centre de Recherches Math\'{e}matiques (CRM), a Natural Sciences and Engineering Research Council of Canada (NSERC) Discovery Grant, and a Canadian Institutes of Health Research (CIHR) grant (CIHR HHP-126781).

We ran computations on the Guillimin and MP2 supercomputers, administered by McGill High-Performance Computing and Universit\'{e} de Sherbrooke, respectively, and managed by Calcul Qu\'{e}bec and Compute Canada. The operation of these supercomputers is funded by the Canada Foundation for Innovation (CFI), minist\`{e}re de l'\'{E}conomie, de la Science et de l'Innovation du Qu\'{e}bec (MESI) and the Fonds de recherche du Qu\'{e}bec - Nature et technologies (FRQ-NT).

The Quebec HIV genotyping program is sponsored by the Minist\`{e}re de la Sant\'{e} et des Services sociaux (MSSS) du Qu\'{e}bec and by the Fonds de recherche du Qu\'{e}bec (FRQ-S) R\'{e}seau SIDA/MI.

\bibliographystyle{chicago} 
\bibliography{clusterAlgoBiblio}

\begin{thebibliography}{}

\bibitem[\protect\citeauthoryear{Ahumada-Ruiz, Flores-Figueroa,
  Toala-González, and Thomson}{Ahumada-Ruiz et~al.}{2009}]{Ahumada-Ruiz2009}
Ahumada-Ruiz, S., D.~Flores-Figueroa, I.~Toala-González, and M.~M. Thomson
  (2009, Sep).
\newblock Analysis of {HIV}-1 pol sequences from {P}anama: identification of
  phylogenetic clusters within subtype {B} and detection of antiretroviral drug
  resistance mutations.
\newblock {\em Infect Genet Evol\/}~{\em 9\/}(5), 933--940.

\bibitem[\protect\citeauthoryear{Bouchard-C{\^{o}}t{\'{e}}, Sankararaman, and
  Jordan}{Bouchard-C{\^{o}}t{\'{e}} et~al.}{2012}]{Bouchard2012}
Bouchard-C{\^{o}}t{\'{e}}, A., S.~Sankararaman, and M.~I. Jordan (2012, Jul).
\newblock Phylogenetic inference via sequential {M}onte {C}arlo.
\newblock {\em Syst Biol\/}~{\em 61\/}(4), 579--593.

\bibitem[\protect\citeauthoryear{Brenner, Wainberg, and Roger}{Brenner
  et~al.}{2013}]{Brenner2013a}
Brenner, B., M.~A. Wainberg, and M.~Roger (2013, Apr).
\newblock Phylogenetic inferences on {HIV}-1 transmission: implications for the
  design of prevention and treatment interventions.
\newblock {\em AIDS\/}~{\em 27\/}(7), 1045--1057.

\bibitem[\protect\citeauthoryear{Brenner, Roger, Routy, Moisi, Ntemgwa, Matte,
  Baril, Thomas, Rouleau, Bruneau, Leblanc, Legault, Tremblay, Charest,
  Wainberg, and {Quebec Primary HIV Infection Study Group}}{Brenner
  et~al.}{2007}]{Brenner2007}
Brenner, B.~G., M.~Roger, J.-P. Routy, D.~Moisi, M.~Ntemgwa, C.~Matte, J.-G.
  Baril, R.~Thomas, D.~Rouleau, J.~Bruneau, R.~Leblanc, M.~Legault,
  C.~Tremblay, H.~Charest, M.~A. Wainberg, and {Quebec Primary HIV Infection
  Study Group} (2007, Apr).
\newblock High rates of forward transmission events after acute/early {HIV}-1
  infection.
\newblock {\em J Infect Dis\/}~{\em 195\/}(7), 951--959.

\bibitem[\protect\citeauthoryear{Brenner, Roger, Stephens, Moisi, Hardy,
  Weinberg, Turgel, Charest, Koopman, Wainberg, and {Montreal PHI Cohort Study
  Group}}{Brenner et~al.}{2011}]{Brenner2011a}
Brenner, B.~G., M.~Roger, D.~A. Stephens, D.~Moisi, I.~Hardy, J.~Weinberg,
  R.~Turgel, H.~Charest, J.~Koopman, M.~A. Wainberg, and {Montreal PHI Cohort
  Study Group} (2011, Oct).
\newblock Transmission clustering drives the onward spread of the {HIV}
  epidemic among men who have sex with men in {Q}uebec.
\newblock {\em J Infect Dis\/}~{\em 204\/}(7), 1115--1119.

\bibitem[\protect\citeauthoryear{Brenner and Wainberg}{Brenner and
  Wainberg}{2013}]{Brenner2013}
Brenner, B.~G. and M.~A. Wainberg (2013, Jul).
\newblock Future of phylogeny in {HIV} prevention.
\newblock {\em J Acquir Immune Defic Syndr\/}~{\em 63 Suppl 2}, S248--S254.

\bibitem[\protect\citeauthoryear{Bryant}{Bryant}{2003}]{Bryant2003}
Bryant, D. (2003).
\newblock A classification of consensus methods for phylogenetics.
\newblock {\em {DIMACS} series in discrete mathematics and theoretical computer
  science\/}~{\em 61}, 163--184.

\bibitem[\protect\citeauthoryear{Chaix, Descamps, Harzic, Schneider, Deveau,
  Tamalet, Pellegrin, Izopet, Ruffault, Masquelier, Meyer, Rouzioux,
  Brun-Vezinet, and Costagliola}{Chaix et~al.}{2003}]{Chaix2003}
Chaix, M.-L., D.~Descamps, M.~Harzic, V.~Schneider, C.~Deveau, C.~Tamalet,
  I.~Pellegrin, J.~Izopet, A.~Ruffault, B.~Masquelier, L.~Meyer, C.~Rouzioux,
  F.~Brun-Vezinet, and D.~Costagliola (2003, Dec).
\newblock Stable prevalence of genotypic drug resistance mutations but increase
  in non-{B} virus among patients with primary {HIV}-1 infection in {F}rance.
\newblock {\em {AIDS}\/}~{\em 17\/}(18), 2635--2643.

\bibitem[\protect\citeauthoryear{Chalmet, Staelens, Blot, Dinakis, Pelgrom,
  Plum, Vogelaers, Vandekerckhove, and Verhofstede}{Chalmet
  et~al.}{2010}]{Chalmet2010}
Chalmet, K., D.~Staelens, S.~Blot, S.~Dinakis, J.~Pelgrom, J.~Plum,
  D.~Vogelaers, L.~Vandekerckhove, and C.~Verhofstede (2010).
\newblock Epidemiological study of phylogenetic transmission clusters in a
  local {HIV}-1 epidemic reveals distinct differences between subtype {B} and
  non-{B} infections.
\newblock {\em BMC Infect Dis\/}~{\em 10}, 262.

\bibitem[\protect\citeauthoryear{Cheon and Liang}{Cheon and
  Liang}{2008}]{Cheon2008}
Cheon, S. and F.~Liang (2008, Jan).
\newblock Phylogenetic tree construction using sequential stochastic
  approximation {M}onte {C}arlo.
\newblock {\em Biosystems\/}~{\em 91\/}(1), 94--107.

\bibitem[\protect\citeauthoryear{Drummond, Suchard, Xie, and Rambaut}{Drummond
  et~al.}{2012}]{Drummond2012}
Drummond, A.~J., M.~A. Suchard, D.~Xie, and A.~Rambaut (2012, Aug).
\newblock Bayesian phylogenetics with {BEAUti} and the {BEAST} 1.7.
\newblock {\em Mol Biol Evol\/}~{\em 29\/}(8), 1969--1973.

\bibitem[\protect\citeauthoryear{Dunn}{Dunn}{1973}]{Dunn1973}
Dunn, J.~C. (1973).
\newblock A fuzzy relative of the {ISODATA} process and its use in detecting
  compact well-separated clusters.
\newblock {\em Journal of Cybernetics\/}.

\bibitem[\protect\citeauthoryear{Eddelbuettel}{Eddelbuettel}{2013}]{Eddelbuettel2013}
Eddelbuettel, D. (2013).
\newblock {\em Seamless {R} and {C++} Integration with {Rcpp}}.
\newblock New York: Springer.
\newblock ISBN 978-1-4614-6867-7.

\bibitem[\protect\citeauthoryear{Eddelbuettel and Fran\c{c}ois}{Eddelbuettel
  and Fran\c{c}ois}{2011}]{Eddelbuettel2011}
Eddelbuettel, D. and R.~Fran\c{c}ois (2011).
\newblock {Rcpp}: Seamless {R} and {C++} integration.
\newblock {\em Journal of Statistical Software\/}~{\em 40\/}(8), 1--18.

\bibitem[\protect\citeauthoryear{Erixon, Svennblad, Britton, and
  Oxelman}{Erixon et~al.}{2003}]{Erixon2003}
Erixon, P., B.~Svennblad, T.~Britton, and B.~Oxelman (2003).
\newblock Reliability of {B}ayesian posterior probabilities and bootstrap
  frequencies in phylogenetics.
\newblock {\em Syst Biol\/}~{\em 52\/}(5), pp. 665--673.

\bibitem[\protect\citeauthoryear{Felsenstein}{Felsenstein}{}]{Felsenstein1981}
Felsenstein, J.
\newblock Evolutionary trees from dna sequences: A maximum likelihood approach.
\newblock {\em Journal of Molecular Evolution\/}~{\em 17\/}(6), 368--376.

\bibitem[\protect\citeauthoryear{Foley, Leitner, Korber, Apetrei, Hahn,
  Mizrachi, Mullins, Rambaut, and Wolinsky}{Foley et~al.}{2013}]{Foley2013}
Foley, B.~T., T.~K. Leitner, B.~T.~M. Korber, C.~Apetrei, B.~Hahn, I.~Mizrachi,
  J.~Mullins, A.~Rambaut, and S.~Wolinsky (2013).
\newblock {HIV} sequence compendium 2013.

\bibitem[\protect\citeauthoryear{Hasegawa, Kishino, and Yano}{Hasegawa
  et~al.}{1985}]{Hasegawa1985}
Hasegawa, M., H.~Kishino, and T.~Yano (1985).
\newblock Dating of the human-ape splitting by a molecular clock of
  mitochondrial {DNA}.
\newblock {\em J Mol Evol\/}~{\em 22\/}(2), 160--174.

\bibitem[\protect\citeauthoryear{Hastings}{Hastings}{1970}]{Hastings1970}
Hastings, W.~K. (1970).
\newblock {M}onte {C}arlo sampling methods using {M}arkov chains and their
  applications.
\newblock {\em Biometrika\/}~{\em 57\/}(1), 97--109.

\bibitem[\protect\citeauthoryear{Holder, Sukumaran, and Lewis}{Holder
  et~al.}{2008}]{Holder2008}
Holder, M.~T., J.~Sukumaran, and P.~O. Lewis (2008).
\newblock A justification for reporting the majority-rule consensus tree in
  {B}ayesian phylogenetics.
\newblock {\em Systematic Biology\/}~{\em 57\/}(5), 814.

\bibitem[\protect\citeauthoryear{Huerta-Cepas, Capella-Guti{\'{e}}rrez,
  Pryszcz, Marcet-Houben, and Gabald{\'{o}}n}{Huerta-Cepas
  et~al.}{2014}]{Huerta-Cepas2014}
Huerta-Cepas, J., S.~Capella-Guti{\'{e}}rrez, L.~P. Pryszcz, M.~Marcet-Houben,
  and T.~Gabald{\'{o}}n (2014, Jan).
\newblock {PhylomeDB} v4: zooming into the plurality of evolutionary histories
  of a genome.
\newblock {\em Nucleic Acids Res\/}~{\em 42\/}(Database issue), D897--D902.

\bibitem[\protect\citeauthoryear{Ibe, Hattori, Fujisaki, Shigemi, Fujisaki,
  Shimizu, Nakamura, Kazumi, Yokomaku, Mamiya, Hamaguchi, and Kaneda}{Ibe
  et~al.}{2008}]{Ibe2008}
Ibe, S., J.~Hattori, S.~Fujisaki, U.~Shigemi, S.~Fujisaki, K.~Shimizu,
  K.~Nakamura, T.~Kazumi, Y.~Yokomaku, N.~Mamiya, M.~Hamaguchi, and T.~Kaneda
  (2008, Jan).
\newblock Trend of drug-resistant {HIV} type 1 emergence among therapy-naive
  patients in {N}agoya, {J}apan: an 8-year surveillance from 1999 to 2006.
\newblock {\em {AIDS} Res Hum Retroviruses\/}~{\em 24\/}(1), 7--14.

\bibitem[\protect\citeauthoryear{Jukes and Cantor}{Jukes and
  Cantor}{1969}]{Jukes1969}
Jukes, T.~H. and C.~R. Cantor (1969).
\newblock Evolution of protein molecules.
\newblock {\em Mammalian protein metabolism\/}~{\em 3\/}(21), 132.

\bibitem[\protect\citeauthoryear{Kimura}{Kimura}{1980}]{Kimura1980}
Kimura, M. (1980, Dec).
\newblock A simple method for estimating evolutionary rates of base
  substitutions through comparative studies of nucleotide sequences.
\newblock {\em J Mol Evol\/}~{\em 16\/}(2), 111--120.

\bibitem[\protect\citeauthoryear{Kouyos, {von Wyl}, Yerly, B{\"{o}}ni, Rieder,
  Joos, Taff{\'{e}}, Shah, B{\"{u}}rgisser, Klimkait, Weber, Hirschel,
  Cavassini, Rauch, Battegay, Vernazza, Bernasconi, Ledergerber, Bonhoeffer,
  G{\"{u}}nthard, and {Swiss HIV Cohort Study}}{Kouyos
  et~al.}{2011}]{Kouyos2011}
Kouyos, R.~D., V.~{von Wyl}, S.~Yerly, J.~B{\"{o}}ni, P.~Rieder, B.~Joos,
  P.~Taff{\'{e}}, C.~Shah, P.~B{\"{u}}rgisser, T.~Klimkait, R.~Weber,
  B.~Hirschel, M.~Cavassini, A.~Rauch, M.~Battegay, P.~L. Vernazza,
  E.~Bernasconi, B.~Ledergerber, S.~Bonhoeffer, H.~F. G{\"{u}}nthard, and
  {Swiss HIV Cohort Study} (2011, Feb).
\newblock Ambiguous nucleotide calls from population-based sequencing of
  {HIV}-1 are a marker for viral diversity and the age of infection.
\newblock {\em Clin Infect Dis\/}~{\em 52\/}(4), 532--539.

\bibitem[\protect\citeauthoryear{Larget and Simon}{Larget and
  Simon}{1999}]{Larget1999}
Larget, B. and D.~L. Simon (1999).
\newblock {M}arkov chain {M}onte {C}arlo algorithms for the {B}ayesian analysis
  of phylogenetic trees.
\newblock {\em Molecular Biology and Evolution\/}~{\em 16}, 750--759.

\bibitem[\protect\citeauthoryear{{Leigh Brown}, Lycett, Weinert, Hughes,
  Fearnhill, Dunn, and {the UK HIV Drug Resistance Collaboration}}{{Leigh
  Brown} et~al.}{2011}]{LeighBrown2011}
{Leigh Brown}, A.~J., S.~J. Lycett, L.~Weinert, G.~J. Hughes, E.~Fearnhill,
  D.~T. Dunn, and {the UK HIV Drug Resistance Collaboration} (2011, Nov).
\newblock Transmission network parameters estimated from {HIV} sequences for a
  nationwide epidemic.
\newblock {\em J Infect Dis\/}~{\em 204\/}(9), 1463--1469.

\bibitem[\protect\citeauthoryear{Lindstr\"{o}m, Ohlis, Huigen, Nijhuis,
  Berglund, Bratt, Sandstr\"{o}m, and Albert}{Lindstr\"{o}m
  et~al.}{2006}]{Lindstroem2006}
Lindstr\"{o}m, A., A.~Ohlis, M.~Huigen, M.~Nijhuis, T.~Berglund, G.~Bratt,
  E.~Sandstr\"{o}m, and J.~Albert (2006).
\newblock {HIV}-1 transmission cluster with {M41L} 'singleton' mutation and
  decreased transmission of resistance in newly diagnosed {S}wedish homosexual
  men.
\newblock {\em Antivir Ther\/}~{\em 11\/}(8), 1031--1039.

\bibitem[\protect\citeauthoryear{Makarenkov, Boc, Xie, Peres-Neto, Lapointe,
  and Legendre}{Makarenkov et~al.}{2010}]{Makarenkov2010}
Makarenkov, V., A.~Boc, J.~Xie, P.~Peres-Neto, F.-J. Lapointe, and P.~Legendre
  (2010).
\newblock Weighted bootstrapping: a correction method for assessing the
  robustness of phylogenetic trees.
\newblock {\em BMC Evol Biol\/}~{\em 10}, 250.

\bibitem[\protect\citeauthoryear{{Pons} and {Latapy}}{{Pons} and
  {Latapy}}{2005}]{Pons2005}
{Pons}, P. and M.~{Latapy} (2005, December).
\newblock {Computing communities in large networks using random walks}.
\newblock {\em ArXiv Physics e-prints\/}.

\bibitem[\protect\citeauthoryear{Posada and Crandall}{Posada and
  Crandall}{2001}]{Posada2001}
Posada, D. and K.~A. Crandall (2001, Jun).
\newblock Selecting models of nucleotide substitution: an application to human
  immunodeficiency virus 1 ({HIV}-1).
\newblock {\em Mol Biol Evol\/}~{\em 18\/}(6), 897--906.

\bibitem[\protect\citeauthoryear{Price, Dehal, and Arkin}{Price
  et~al.}{2010}]{Price2010}
Price, M.~N., P.~S. Dehal, and A.~P. Arkin (2010, 03).
\newblock {FastTree} 2 – approximately maximum-likelihood trees for large
  alignments.
\newblock {\em PLOS ONE\/}~{\em 5\/}(3), 1--10.

\bibitem[\protect\citeauthoryear{Prosperi, Ciccozzi, Fanti, Saladini, Pecorari,
  Borghi, Giambenedetto, Bruzzone, Capetti, Vivarelli, Rusconi, Re, Gismondo,
  Sighinolfi, Gray, Salemi, Zazzi, Luca, and {on behalf of the ARCA
  collaborative group}}{Prosperi et~al.}{2011}]{Prosperi2011}
Prosperi, M. C.~F., M.~Ciccozzi, I.~Fanti, F.~Saladini, M.~Pecorari, V.~Borghi,
  S.~D. Giambenedetto, B.~Bruzzone, A.~Capetti, A.~Vivarelli, S.~Rusconi, M.~C.
  Re, M.~R. Gismondo, L.~Sighinolfi, R.~R. Gray, M.~Salemi, M.~Zazzi, A.~D.
  Luca, and {on behalf of the ARCA collaborative group} (2011, May).
\newblock A novel methodology for large-scale phylogeny partition.
\newblock {\em Nat Commun\/}~{\em 2}, 321.

\bibitem[\protect\citeauthoryear{Ragonnet-Cronin, Hodcroft, Hu{\'e}, Fearnhill,
  Delpech, Brown, and Lycett}{Ragonnet-Cronin
  et~al.}{2013}]{Ragonnet-Cronin2013}
Ragonnet-Cronin, M., E.~Hodcroft, S.~Hu{\'e}, E.~Fearnhill, V.~Delpech,
  A.~J.~L. Brown, and S.~Lycett (2013, Nov).
\newblock Automated analysis of phylogenetic clusters.
\newblock {\em BMC Bioinformatics\/}~{\em 14\/}(1), 317.

\bibitem[\protect\citeauthoryear{Revell}{Revell}{2012}]{Revell2012}
Revell, L.~J. (2012).
\newblock phytools: An {R} package for phylogenetic comparative biology (and
  other things).
\newblock {\em Methods in Ecology and Evolution\/}~{\em 3}, 217--223.

\bibitem[\protect\citeauthoryear{Ronquist, Teslenko, van~der Mark, Ayres,
  Darling, Höhna, Larget, Liu, Suchard, and Huelsenbeck}{Ronquist
  et~al.}{2012}]{Ronquist2012}
Ronquist, F., M.~Teslenko, P.~van~der Mark, D.~L. Ayres, A.~Darling, S.~Höhna,
  B.~Larget, L.~Liu, M.~A. Suchard, and J.~P. Huelsenbeck (2012).
\newblock {MrBayes} 3.2: Efficient {Bayesian} phylogenetic inference and model
  choice across a large model space.
\newblock {\em Systematic Biology\/}~{\em 61\/}(3), 539--542.

\bibitem[\protect\citeauthoryear{Sanderson and Curtin}{Sanderson and
  Curtin}{2016}]{Sanderson2016}
Sanderson, C. and R.~Curtin (2016).
\newblock Armadillo: a template-based {C++} library for linear algebra.
\newblock {\em Journal of Open Source Software\/}~{\em 1\/}(2), 26--32.

\bibitem[\protect\citeauthoryear{Schliep}{Schliep}{2011}]{Schliep2011}
Schliep, K. (2011).
\newblock phangorn: phylogenetic analysis in {R}.
\newblock {\em Bioinformatics\/}~{\em 27\/}(4), 592--593.

\bibitem[\protect\citeauthoryear{Stamatakis}{Stamatakis}{2014}]{Stamatakis2014}
Stamatakis, A. (2014).
\newblock {RAxML} version 8: A tool for phylogenetic analysis and post-analysis
  of large phylogenies.
\newblock {\em Bioinformatics\/}.

\bibitem[\protect\citeauthoryear{Stamatakis, Ludwig, and Meier}{Stamatakis
  et~al.}{2005}]{Stamatakis2005}
Stamatakis, A., T.~Ludwig, and H.~Meier (2005).
\newblock {RAxML-III}: a fast program for maximum likelihood-based inference of
  large phylogenetic trees.
\newblock {\em Bioinformatics\/}~{\em 21\/}(4), 456.

\bibitem[\protect\citeauthoryear{Susko}{Susko}{2009}]{Susko2009}
Susko, E. (2009, Apr).
\newblock Bootstrap support is not first-order correct.
\newblock {\em Syst Biol\/}~{\em 58\/}(2), 211--223.

\bibitem[\protect\citeauthoryear{Swofford}{Swofford}{2003}]{Swofford2003}
Swofford, D.~L. (2003).
\newblock {PAUP*}: Phylogenetic analysis using parsimony (and other methods).

\bibitem[\protect\citeauthoryear{Tamura, Stecher, Peterson, Filipski, and
  Kumar}{Tamura et~al.}{2013}]{Tamura2013}
Tamura, K., G.~Stecher, D.~Peterson, A.~Filipski, and S.~Kumar (2013, Dec).
\newblock {MEGA6}: Molecular evolutionary genetics analysis version 6.0.
\newblock {\em Mol Biol Evol\/}~{\em 30\/}(12), 2725--2729.

\bibitem[\protect\citeauthoryear{{Van der Spoel van Dijk}, Makhoahle, Rigouts,
  and Baba}{{Van der Spoel van Dijk} et~al.}{2016}]{VanderSpoelvanDijk2016}
{Van der Spoel van Dijk}, A., P.~M. Makhoahle, L.~Rigouts, and K.~Baba (2016).
\newblock Diverse molecular genotypes of {M}ycobacterium tuberculosis complex
  isolates circulating in the {F}ree {S}tate, {S}outh {A}frica.
\newblock {\em Int J Microbiol\/}~{\em 2016}, 6572165.

\bibitem[\protect\citeauthoryear{Villandre, Stephens, Labbe, G{\"{u}}nthard,
  Kouyos, Stadler, and {Swiss HIV Cohort Study}}{Villandre
  et~al.}{2016}]{Villandre2016}
Villandre, L., D.~A. Stephens, A.~Labbe, H.~F. G{\"{u}}nthard, R.~Kouyos,
  T.~Stadler, and {Swiss HIV Cohort Study} (2016).
\newblock Assessment of overlap of phylogenetic transmission clusters and
  communities in simple sexual contact networks: Applications to {HIV}-1.
\newblock {\em PLoS One\/}~{\em 11\/}(2), e0148459.

\bibitem[\protect\citeauthoryear{Vrbik, Stephens, Roger, and Brenner}{Vrbik
  et~al.}{2015}]{Vrbik2015}
Vrbik, I., D.~A. Stephens, M.~Roger, and B.~G. Brenner (2015).
\newblock The {G}ap procedure: for the identification of phylogenetic clusters
  in {HIV}-1 sequence data.
\newblock {\em BMC Bioinformatics\/}~{\em 16}, 355.

\bibitem[\protect\citeauthoryear{Wang and Yang}{Wang and Yang}{2014}]{Wang2014}
Wang, Y. and Z.~Yang (2014).
\newblock Priors in {B}ayesian phylogenetics.
\newblock {\em {B}ayesian phylogenetics: methods, algorithms, and applications.
  {C}hapman and {H}all/{CRC}\/}, 5--23.

\bibitem[\protect\citeauthoryear{Yang}{Yang}{2006}]{Yang2006}
Yang, Z. (2006).
\newblock {\em Computational Molecular Evolution}.
\newblock Oxford Series in Ecology and Evolution. Oxford: Oxford University
  Press.

\bibitem[\protect\citeauthoryear{Yang and Rannala}{Yang and
  Rannala}{1997}]{Yang1997}
Yang, Z. and B.~Rannala (1997, Jul).
\newblock Bayesian phylogenetic inference using {DNA} sequences: a {Markov}
  {Chain} {Monte} {Carlo} method.
\newblock {\em Mol Biol Evol\/}~{\em 14\/}(7), 717--724.

\end{thebibliography}

\section*{Supplementary Material S1 - Algorithm description}

\textbf{Input}: 
\begin{enumerate}
 \item \textbf{Topology}: Can be, for example, the maximum likelihood topology,
 \item \textbf{Nucleotide transition rate matrix}: Can be an empirical estimate, like the one in \cite{Posada2001}, or alternatively, one derived from the sample itself, with the help of RAxML or MrBayes for example,
 \item \textbf{Gamma shape parameter for among-loci mutation rate variation}: Assumed equal to the scale parameter, can be obtained in the same way as the nucleotide transition rate matrix. In the simulations, we use an estimate from \cite{Posada2001},  
 \item \textbf{Cluster membership indices prior}: Follows a Dirichlet-multinomial distribution, combined with a Poisson-distributed weight with a pre-determined rate parameter, e.g. the number of clusters resulting from a conventional bootstrap-maximum likelihood phylogenetic clustering analysis,
 \item \textbf{Poisson rate for the assumed number of clusters},
 \item \textbf{Concentration parameter prior}: Assumed gamma-distributed with user-specified scale and shape parameters,
 \item \textbf{Shape and scale parameter values for the concentration parameter prior}: We set the scale parameter equal to $0.1$ in all analyses, and changed the shape parameter to vary the distributional mean,
 \item \textbf{Transition kernel for the concentration parameter}: A uniform distribution with radius $0.5$ centered at the current parameter value,
 \item \textbf{Transition kernel for the cluster membership indices}: 
 A uniform distribution over all configurations reachable from the current state. A configuration is reachable if it can be obtained by splitting in two a cluster of size $2$ or more, or merging two neighbouring clusters. Two clusters are considered neighbours if their respective most recent common ancestors (MRCA) are siblings. Clusters are obtained by partitioning the sample into disjoint clades. It follows that each cluster can be represented, alternatively, by its MRCA. When a cluster is split in two, the MRCAs of the new clusters are the children nodes of the original cluster's MRCA. When two neighbouring clusters are merged, the new cluster's MRCA is the parent node of the selected two clusters' MRCAs.
 \item \textbf{Prior for branch lengths in the within-cluster phylogenies}: Assumed to follow the exponential distribution,
 \item \textbf{Prior for branch lengths in the within-cluster phylogeny}: Assumed to follow a log-normal distribution with equal mean and standard deviation, which implies a coefficient of variation of $1$,
 \item \textbf{Prior for the transition probabilities along branches in the within-cluster phylogenies}: 
 Represented by an array of $4\times4$ matrices. Each row of the array corresponds to a different assumed mean branch length, while each column corresponds to a different rate variation category,  
 \item \textbf{Prior for the transition probabilities along branches in the between-cluster phylogeny}:
 Same as before,
 \item \textbf{Starting value for the cluster membership indices}: Must be a partition of the sample into clades found in the input topology, 
 \item \textbf{Starting value for the Dirichlet-multinomial concentration parameter},
 \item \textbf{Starting values for the between-cluster and within-cluster transition probabilities},
 \item \textbf{Number of iterations},
 \item \textbf{Burn-in size},
 \item \textbf{Thinning ratio}.
\end{enumerate}

\textbf{Algorithm output}:
\begin{enumerate}
 \item Values sampled from the posterior distribution of the cluster membership indices,
 \item Values sampled from the posterior distribution of the concentration parameter,
 \item A non-standardized joint log-posterior probability value for the parameter values at the end of each iteration.
\end{enumerate}

\subsection*{A standard run}

\subsubsection*{Obtaining the topology}

In each simulation run, we start by obtaining an estimate of the maximum likelihood topology from RAxML. We assume that genetic distances follow the GTR+$\Gamma(5)$ model and use a subtype C outgroup (\url{http://www.hiv.lanl.gov/}, accession number: AB254141). We then produce $500$ bootstrap estimates of the tree, resulting in the usual clade support estimates. RAxML stores the best scoring tree in a file with the ``bestTree'' mention. More details on RAxML's tree optimization and scoring methods can be found in \cite{Stamatakis2005}.  

\subsubsection*{Starting values for the cluster membership indices}

We then use the topology to obtain initial cluster estimates. More specifically, we look for a partition of the sample into clades for which,
\begin{enumerate}
\item Maximum patristic distance between any pair of elements within a clade is bounded above by an arbitrary value, e.g. $5$\%,
\item Bootstrap support for any clade is above a certain value, e.g. $70$\%. 
\end{enumerate}
We find such a partition by traversing the tree starting at the root. At the beginning, all sequences are assumed to be in one cluster. If the (trivial) clade supported by the root node meets the requirements above, no further move is required. If not, we move down to the two children nodes, and update the cluster membership vector to account for the creation of a new cluster after the split of the original cluster into two non-overlapping clusters. At each child, we repeat the checks performed at the root, moving down and splitting clusters until a set that meets the clustering criteria is encountered, or until we reach a tip.

In the analyses, we impose a confidence requirement of $70$\%, and find cluster configurations for maximum genetic distance requirements between $3$\% and $12$\%. For each distance requirement, we have a potentially different set of clusters, and for each of them, we calculate the Dunn index \cite{Dunn1973}, deriving the distance matrix from the phylogenetic estimate. Finally, we pick the set that maximizes that index as the starting value for the cluster membership indices.

\subsubsection*{Estimates of transition probabilities}

Once we have an estimate of cluster membership indices, we use it to set up priors for transition probabilities along branches in the within-cluster and between-cluster phylogenies. In the within-cluster phylogenies, branch lengths have an exponential prior. We pick a range of values for the mean parameter by,
\begin{enumerate}
 \item Computing the average branch length across all within-cluster phylogenies obtained from the starting partition,
 \item Finding $20$ equidistant points in a radius equal to $8$\% of the value computed previously.
\end{enumerate}
For each point in the range, we simulate $100,000$ values from the corresponding exponential distribution. We then obtain the required transition probability matrices by computing,
\begin{equation*}
 P^{(r)} = \sum_{i=1}^{1e5} \exp(Qd_i l_r)/1e5, \hspace{0.5cm} r = 1,2,3,
\end{equation*}
where $r$ indexes the rate variation category, $d_i$ denotes a value generated previously, $Q$, a transition rate matrix estimate, and $l_r$, a distance scaling factor. We use a similar strategy to derive a prior distribution for transition probabilities along branches in the between-cluster phylogeny. 

\subsubsection*{Running the chain and obtaining point estimates for cluster membership indices}

Each iteration in the chain involves successive Metropolis-Hastings updates of the cluster membership indices, the between and within-cluster transition probabilities, and the concentration parameter. The algorithm produces a joint posterior probability value at the end of each iteration, which we use to identify the MAP estimate. To obtain the linkage-$xx$ estimates, we compute an adjacency matrix from each sampled cluster membership vector, under the assumption that all sets of co-clustering sequences form fully-connected graphs, all disjoint from each other. We then average all adjacency matrices, and apply the $xx$ threshold to the resulting matrix, rounding up to $1$ all values in the matrix above the threshold, and down to $0$ the other values. We then run the walktrap algorithm \cite{Pons2005}, using chains of $10$ steps to detect disjoint sets, which correspond to the cluster membership indices estimate.

\section*{Supplementary Material S2 - Tuning parameters used in the simulations}

\subsection*{Simulating datasets}

\begin{itemize}
 \item Sample size: $200$,
 \item Rate parameter for Poisson-distributed number of clusters: $50$,
 \item Mean value for normally-distributed concentration parameter: $10$,
 \item Standard deviation for normally-distributed concentration parameter: $2$,
 \item Number of rate variation categories: $5$,
 \item Shape and scale parameters for gamma-distributed rate variation: $0.7589$,
 \item Number of datasets: $100$,
 \item Root sequence: HXB2 sequence (\url{http://www.hiv.lanl.gov/}), sites $10$-$297$ of the protease region (PR), and $112$-$741$ of the reverse transcriptase (RT) region, of the \textit{pol} gene.
 \item Limiting probabilities: $(A = 0.39, T = 0.22, C = 0.17, G = 0.22)$
 \item Rate matrix $Q$:
 $$
 \begin{bmatrix} 
    -0.83708096 & 0.04319486 & 0.12127074 & 0.67261536 \\
    0.07657272 & -0.82554421 & 0.66140131 & 0.08757018 \\
    0.27820934 & 0.85593111 & -1.18569748 & 0.05155703 \\
    1.19236359 & 0.08757018 & 0.03983952 & -1.31977330
 \end{bmatrix}
 $$
  \item Mean parameter for exponentially-distributed branch lengths in within-cluster phylogenies: $0.003$,
  \item Mean and standard deviation parameters for log-normal-distributed branch lengths in between-cluster phylogenies: $0.008$.
\end{itemize}

\subsection*{Chain parameters}

\begin{itemize}
 \item Number of discrete states for the within-cluster and between-cluster transition probability matrices: $20$,
 \item Number of samples used to obtain transition probability matrices: $100,000$,
 \item Radius around mean within-cluster and between-cluster branch length estimates: $8$\%,
 \item Bootstrap confidence requirement for initial cluster estimate: $70$\%,
 \item Limiting probabilities: $(A = 0.4298969, T = 0.2227602, C = 0.1459, G = 0.2014428)$,
 \item Rate matrix Q:
 $$
 \begin{bmatrix}
    -0.79633415 & 0.04560603 & 0.10852696 & 0.64220116 \\
    0.08801344 & -0.76352160 & 0.59189771 & 0.08361045 \\
    0.31977658 & 0.90370975 & -1.27271206 & 0.04922573 \\
    1.37051455 & 0.09245841 & 0.03565297 & -1.49862593
 \end{bmatrix}
 $$
  \item Shape parameter for concentration parameter prior: $1000$, $100$, $10$,
  \item Scale parameter for concentration parameter prior: $0.1$,
  \item Poisson rate for weight applied to the cluster membership vector prior: $50$,
  \item Number of iterations: $55,000$.
\end{itemize}

\section*{Supplementary Material S3 - Tuning parameters used in the real data analysis}

\subsection*{Bootstrap analysis}

\begin{itemize}
 \item Number of discrete states for the within-cluster and between-cluster transition probability matrices: $20$,
 \item Number of samples used to obtain transition probability matrices: $100,000$,
 \item Radius around mean within-cluster and between-cluster branch length estimates: $8$\%,
 \item Discrete gamma distribution parameter: $0.7589$,
 \item Bootstrap confidence requirement for initial cluster estimate: $70$\%,
 \item Limiting probabilities: $(A = 0.39, T = 0.22, C = 0.17, G = 0.22)$,
 \item Rate matrix $Q$:
 $$
 \begin{bmatrix} 
    -0.83708096 & 0.04319486 & 0.12127074 & 0.67261536 \\
    0.07657272 & -0.82554421 & 0.66140131 & 0.08757018 \\
    0.27820934 & 0.85593111 & -1.18569748 & 0.05155703 \\
    1.19236359 & 0.08757018 & 0.03983952 & -1.31977330
 \end{bmatrix}
 $$ 
 \item Shape parameter for concentration parameter prior: $1000$,
 \item Scale parameter for concentration parameter prior: $0.1$,
 \item Poisson rate for weight applied to the cluster membership vector prior: $32$,
 \item Number of iterations: $55,000$.
\end{itemize}

\subsection*{Approximation of the fully Bayesian analysis}

\begin{itemize}
 \item Number of discrete states for the within-cluster and between-cluster transition probability matrices: $20$,
 \item Number of samples used to obtain transition probability matrices: $100,000$,
 \item Radius around mean within-cluster and between-cluster branch length estimates: $8$\%,
 \item Discrete gamma distribution parameter: $0.4394492$,
 \item Limiting probabilities: $(A = 0.4032267, T = 0.2147781, C = 0.1625374, G = 0.2194578)$,
 \item Rate matrix $Q$:
 $$
 \begin{bmatrix} 
  -0.8411512 &  0.05921394 & 0.11223579 & 0.66970147 \\
  0.1111689 & -0.80528701 & 0.62140549 & 0.07271263 \\
  0.2784372 & 0.82112972 & -1.17182113 & 0.07225417 \\
  1.2304940 & 0.07116212 & 0.05351373 & -1.35516988
 \end{bmatrix}
 $$ 
 \item Shape parameter for concentration parameter prior: $1000$,
 \item Scale parameter for concentration parameter prior: $0.1$,
 \item Poisson rate for weight applied to the cluster membership vector prior: $32$,
 \item Number of iterations: $55,000$.
\end{itemize}

\subsection*{Main run}

\begin{itemize}
 \item Number of discrete states for the within-cluster and between-cluster transition probability matrices: $20$,
 \item Number of samples used to obtain transition probability matrices: $100,000$,
 \item Radius around mean within-cluster and between-cluster branch length estimates: $8$\%,
 \item Discrete gamma distribution parameter: $0.7589$,
 \item Bootstrap confidence requirement for initial cluster estimate: $70$\%,
 \item Limiting probabilities: $(A = 0.39, T = 0.22, C = 0.17, G = 0.22)$,
 \item Rate matrix $Q$:
 $$
 \begin{bmatrix} 
    -0.83708096 & 0.04319486 & 0.12127074 & 0.67261536 \\
    0.07657272 & -0.82554421 & 0.66140131 & 0.08757018 \\
    0.27820934 & 0.85593111 & -1.18569748 & 0.05155703 \\
    1.19236359 & 0.08757018 & 0.03983952 & -1.31977330
 \end{bmatrix}
 $$ 
 \item Shape parameter for concentration parameter prior: $1000$,
 \item Scale parameter for concentration parameter prior: $0.1$,
 \item Poisson rate for weight applied to the cluster membership vector prior: $32$,
 \item Number of iterations: $220,000$.
\end{itemize}

\section*{Supplementary Material S4 - Notes on the software}

We implemented DM-PhyClus mostly in R, with C++ modules to handle log-likelihood evaluations. In R, we use classes and functions defined in the \textit{ape} and \textit{phangorn} packages \cite{Schliep2011} to represent and manipulate phylogenies. The interface between R and C++ relies on features offered by the \textit{Rcpp} and \textit{RcppArmadillo} packages. \cite{Eddelbuettel2011, Eddelbuettel2013}. 

Unsurprisingly, the C++ modules make extensive use of containers in the Standard Template Library (STL) and functionalities implemented in the C++11 standard. For now, the code still relies on the GNU Scientific Library (GSL) for random number generation, but we intend to change that in future versions in order to improve portability. Phylogenies are represented by a custom binary tree class, consisting of objects instanced from an input node class, representing the tips of the tree, and from an internal node class. Both classes inherit from an abstract class, standing in for a generic tree node. 

We use Felsenstein's tree-pruning algorithm \cite{Felsenstein1981} to perform likelihood evaluations. Our implementation of the latter algorithm makes use of containers, functions, and operators defined in the Armadillo library \cite{Sanderson2016}. To reduce the algorithm's memory footprint and improve performance, all intermediate solutions are saved in a map container, and the tree node objects store merely a pointer to the corresponding map elements. To ensure pointer validity, we opted for an ordered map. We use functions in the \textit{boost} package in the generation of keys for map elements. The keys are obtained recursively by combining, among other things, keys computed for children nodes.

The size of the map tends to increase quickly for even moderately-sized datasets, eventually saturating the memory on most standard machines, and so, the software wipes the map periodically. That strategy is also beneficial from a computational standpoint: by eliminating configurations rarely visited by the algorithm, mean lookup time is reduced. Moreover, allowing very large maps is detrimental from a computational standpoint: once a map reaches a certain size, re-computing solutions turns out to be on average faster than doing a lookup.   

We obtained a great boost in performance after defining a persistent pointer to the object used to represent the tree structure. Indeed, profiling had revealed that the software was being weighed down considerably by the memory allocation operations involved in building the tree structure, hence the vast improvement resulting from keeping the object in memory and updating it when required. More specifically, we implemented that strategy by passing a so-called \textit{external pointer} to R, implemented by the XPtr class template in the Rcpp library. By trading the pointer between R and C++, we effectively prevent garbage collection of the tree object until the pointer goes out of scope. 

We wrote a vignette that explains how the R package can be used to cluster an arbitrary dataset. 

\section*{Supplementary Material S5 - Log-posterior probability graph}

\begin{figure}[h!]
  \includegraphics[width = 7cm, angle = -90]{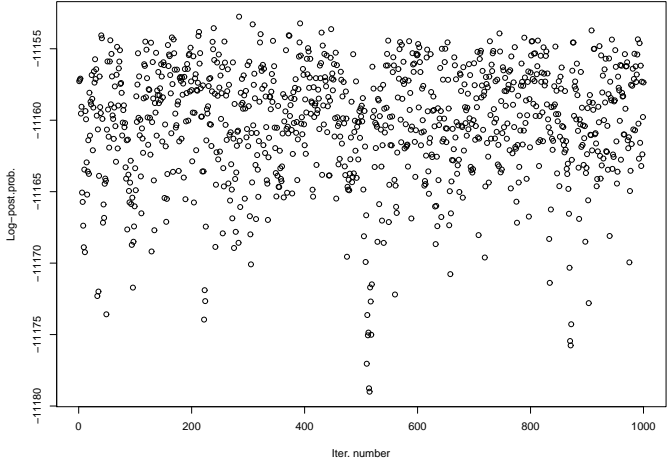}
  \caption[Log-posterior probability graph for the thinned chain obtained from one of the simulated samples]{\textbf{Log-posterior probability graph for the thinned chain obtained from one of the simulated samples.}} \label{fig:logPostProbGraph}
\end{figure}

See Figure \ref{fig:logPostProbGraph}.

\end{document}